\documentclass[%
 reprint,
 amsmath,amssymb,
 aps,
 pre,
]{revtex4-2}
\usepackage{graphicx}
\usepackage{dcolumn}
\usepackage[caption=false]{subfig}
\usepackage{bm}
\usepackage{appendix}

\begin{document}

\preprint{APS/123-QED}

\title{Experimental Tests of Radio-Frequency Heating Saturation in Ultracold Neutral Plasmas}

\author{Bridget O'Mara}
\author{Ryan C. Baker}
\author{Jacob L. Roberts}

\affiliation{Physics Department, Colorado State University, Fort Collins, Colorado 80523, USA}

\date{\today}

\begin{abstract}
For non-resonant radio-frequency (RF) fields, electron heating in sufficiently collisional plasmas can be driven primarily by inverse bremsstrahlung absorption. When the quiver velocity $v_{osc}$ approaches the electron thermal velocity $v_{th}$, theory often predicts sublinear scaling of the heating rate with RF power, indicating saturation. We experimentally test this prediction in ultracold neutral plasmas by finding RF pulses of different amplitude and duration that produce the same electron heating. Despite $v_{osc} \sim v_{th}$, we measured no observable saturation. We compare our results to linear response theory (LRT) and a binary collision theory (BCT). The predicted saturation in both theories is sensitive to how common assumptions about cutoff parameters are applied, and agreement with experimental results is much better if quiver-velocity-dependent cutoffs in LRT and BCT are used. Additionally, under our conditions of moderate coupling and magnetization, we find no evidence that RF heating distorts the electron velocity distribution from Maxwell–Boltzmann, indicating saturation from the Langdon effect is suppressed.  
\end{abstract}

\maketitle
\section{Introduction}
Inverse bremsstrahlung absorption (IBA)~\cite{DawsonOberman1962,Decker1994} is an important, and often the most important, energy transfer mechanism from electromagnetic (EM) radiation to electron temperature in laser-plasma systems~\cite{Wiggins2010,Milder2021} including systems in the warm dense matter regime~\cite{Falk2018}, those with fusion-relevant conditions~\cite{Hurricane2023,Turnbull,Farrashbandi2020b,Oussama2021}, and those with other technological applications such as laser penetration welding~\cite{ZHANG2022139434}. It is also relevant for radio-frequency (RF) heating of lower density plasmas~\cite{Gustavsson2010,Smorenburg2012}. IBA arises from electron–ion collisions in an oscillating electric field that disrupt the phase of the electrons' driven motion and result, on average, in a net increase in electron kinetic energy and thus temperature. At low EM intensities, IBA heating scales linearly with the radiation intensity. With increasing intensity, the scaling becomes sublinear, indicating saturation \cite{Langdon1980,Mulser,Geltman1977}. Saturation is expected when the amplitude of the electron quiver velocity in the EM field, $v_{osc}$, becomes comparable to the electron thermal velocity $v_{th}=\sqrt{k_B T_e/m_e}$, where $T_e$ is the electron temperature, $m_e$ is the electron mass, and $k_B$ is Boltzmann's constant~\cite{Langdon1980,Jones1982,Mulser}. It is attributed to the increase in the mean electron–ion collision velocity with $v_{osc}$, which reduces the Coulomb collision cross section. Lower-velocity electrons therefore receive preferential heating that drives the electron velocity distribution away from a Maxwell-Boltzmann (MB) in a phenomenon known as the Langdon effect~\cite{Langdon1980}. In some theory treatments, saturation and the Langdon effect are expected to be mitigated by increases in effective electron screening length with $v_{osc}$, leading to an increased collision rate~\cite{Mulser,Daligault2009,Gericke2002,Langdon1980}.

Saturation measurements thus probe the intensity dependence of the electron–ion collision rate in a plasma. If no effects occur other than an increase in the mean collision velocity, the IBA would scale as $\left( v_{th}^2+v_{osc}^2/6\right)^{-3/2}$. Deviations from this scaling indicate intensity-dependent changes in screening, deviations of the electron velocity distribution from Maxwell–Boltzmann, or other effects such as correlations caused by multiple collisions between the same electron and ion~\cite{Decker1994}. Our experimental measurements characterize the importance of these effects and, moreover, enable comparison with theoretical predictions, providing data collected under conditions not previously explored.

In experiments reported here, RF pulses were used to heat ultracold neutral plasmas (UNPs)~\cite{Killian1999,Killian2010,Lyon2017}. These UNPs were created by photoionizing an ultracold atomic gas~\cite{GuthrieDens2024}, producing free electrons and ions. After ionization, electrons escaped until the resulting space charge confined the remainder. The ions remained essentially stationary on our experiment timescales. The resulting UNP was a finite, two‑component plasma with moderate density and near‑absolute‑zero temperatures.

The properties of UNPs make them experimental platforms well-suited to study fundamental plasma properties, such as IBA, using a table-top-scale experimental apparatus. Since the RF wavelengths are much larger than the UNPs, IBA measurements are not complicated by optics. 
Because of their cold electron temperatures, plasmas of moderate coupling strength can be measured~\cite{Weiting2017}.
For this initial exploration of RF heating saturation in UNPs, we chose parameters for our apparatus that gave a high degree of sensitivity to any RF saturation. The UNPs were in the moderately coupled regime with an average Coulomb coupling parameter $\Gamma=e^2/(4\pi\epsilon_0ak_BT) = 0.13$~\cite{Ichimaru1982}, where $e$ is the electron charge, $\epsilon_0$ is the vacuum permittivity and $a$ is the Wigner-Seitz radius. To enhance the experimental signal, a magnetic field of $10.7\,\mathrm{G}$ was applied. Under these conditions the UNPs were weakly magnetized~\cite{BaalrudDaligault}, with theory estimates predicting negligible ($<1\%$) changes in heating rate compared to zero magnetic field.

Recent measurements of IBA reported in refs.~\cite{Turnbull,Milder2021} were sensitive to saturation and show agreement with theory predictions only when ion contributions to screening, specific treatments of the electron–ion collision rate, and changes in the electron velocity distribution~\cite{Langdon1980} during heating are taken into account. In contrast, we do not observe any saturation at the few percent level for ratios of $v_{osc}/v_{th}$ up to 1.3. This does not necessarily imply disagreement between measurements since the plasma parameters studied, such as degree of electron strong coupling~\cite{Gericke2002,Dimonte2008} and electron-ion collision rate per RF oscillation~\cite{Cauble1985,Krook1976}, are not the same. But, it does indicate that the saturation effects shown in those prior works are not observed under our experimental conditions. 

IBA is commonly described using linear response theory (LRT)~\cite{DawsonOberman1962}, which treats electrons as a dielectric fluid, and binary collision theory (BCT)~\cite{Mulser}, which treats them as colliding particles. Under standard assumptions, both approaches predict saturation for ratios of $v_{osc}/v_{th}$ we induced, in disagreement with our measurements. Agreement is recovered when fixed cutoffs are replaced with quiver-velocity‑dependent cutoffs (VDCs) in LRT \cite{Grabowski2013} and BCT~\cite{Devriendt2022}. The following sections discuss these different treatments along with what a saturation-free model for meaningful comparisons. Expected impact of the Langdon effect is also considered for our conditions. In the work presented, we compare our measurements with LRT and BCT predictions under these differing assumptions. We describe these predictions (Section~\ref{sec:theory}), then describe our experimental techniques (Section~\ref{sec:ExperimentalApp}) and results (Section~\ref{sec:ResultsandDiscussion}), and then draw conclusions from those results (Section~\ref{sec:Conclusions}).

\section{Theory Predictions of RF Saturation}
\label{sec:theory}

Both LRT and BCT predict IBA saturation for sufficiently large driving field amplitudes. The extent of saturation predicted, however, depends strongly on assumptions made in each theory treatment for our experimental conditions. We describe both LRT and BCT in the next two subsections along with a range of assumptions commonly used in these theories.


\subsection{Linear Response Theory}
In LRT, the electrons are treated as a dielectric fluid. IBA occurs as the ions' motion relative to the electrons produces wakes in the dielectric fluid that in turn lead to forces that ultimately heat the electrons.
Saturation occurs because strong RF fields drive dielectric fluid responses at higher harmonics, where the associated heating is less efficient. Within the LRT model, we connect the energy loss $-dH$ for an ion over a distance $dl$~\cite{DawsonOberman1962,PeterT1991,Maynard1982,bekefi1966radiation}, the RF-induced oscillation of electrons relative to ions in a plasma, the linearized Vlasov-Poisson equations~\cite{PeterT1991,Fitzpatrick2014}, and energy conservation to find a heating rate~\cite{Nersisyan2007book,Guthrie2021}:
\begin{align}
\frac{dH}{dt} &= \frac{2i e^2}{4\pi \epsilon_0 (2\pi)^3}
\int_0^{k_{max}} \mathrm{d}^3 \mathbf{k} \int_{-\infty}^{\infty} \mathrm{d}\omega \,
\frac{\mathbf{k} \cdot \mathbf{\hat{z}}}{k^2 \, \epsilon(\mathbf{k}, \omega)}\frac{\mathrm{d}z(t)}{\mathrm{dt}}
\notag \\
&\quad \times \exp[i \mathbf{k} \cdot \mathbf{\hat{z}}z(t) - i\omega t]
\notag \\
&\times \int_{-\infty}^{\infty} \mathrm{d}t' \,
\exp[-i \mathbf{k} \cdot \mathbf{\hat{z}}z(t') + i \omega t']
\label{eq:bigSVint}
\end{align}
where $z(t)$ is the electron center-of-mass (COM) position (the electron quiver velocity is dz(t)/dt), $\omega$ is the frequency of the plasma's response to a perturbation, and $\mathbf{k}$ is the wavenumber of the plasma response. $\epsilon(\mathbf{k},\omega)$ is the plasma dielectric response function~\cite{Nersisyan2007book,Guthrie2021}, given by,
\begin{equation}
\begin{aligned}
&\epsilon(\mathbf{k}, \omega_r) = \\
&\quad 1 + \frac{1}{k^2 \lambda_D^2} \left[ 
1 + \frac{\omega_r}{\sqrt{2} |k_\parallel| v_{\text{th}}} 
\sum_{m=-\infty}^{\infty} e^{-\beta_k} I_m(\beta_k) Z\left( \frac{x_m}{\sqrt{2}} \right)
\right]
\end{aligned}
\end{equation}
with
\begin{equation}
    \beta=\frac{k_{\perp}^2v_{th}^2}{\omega_c^2}.
\end{equation}
Where $\omega_c$ is the cyclotron frequency, $\omega_r$ is the frequency of the plasma's response to a perturbation, and $k_{||}$ and $k_{\perp}$ denote the wavenumbers of the plasma response parallel and orthogonal to the magnetic field respectively. $I_m(\beta_k)$ is the $m^{th}$ order modified Bessel function of the 1st kind, and $Z(x_m/\sqrt{2})$ is the plasma dispersion function~\cite{FriedConte1961}. 
This is the dielectric function associated with a collisionless, magnetized plasma~\cite{Nersisyan2007book,Lafleur2020,Krall1969}. For our conditions, collisions are not expected to have a significant effect on the LRT predictions for heating rates Ref.~\cite{Guthrie2021} and even less of an effect on the predicted saturation of those rates. 

$k_{max}$ is a cutoff parameter necessary to avoid logarithmic divergence, set to $1/r_{min}$, where $r_{min}$ can be thought of as the impact parameter limit for high-angle deflections. This definition for $r_{min}$ is used because LRT overcounts the heating contribution from small-impact parameter, large-angle deflections~\cite{DawsonOberman1962,bekefi1966radiation,PeterT1991}. A common choice for $r_{min}$ is the Landau length, $r_{L}=e^2/(4\pi\epsilon_0m_e v _{th}^2)$~\cite{PeterT1991,Grabowski2013,DawsonOberman1962,Nersisyan2007book}. In the context of stopping power, Ref.~\cite{Grabowski2013} showed molecular dynamics simulations were consistent with a value of $r_{min}$ that included a dependence on the ion velocity through the plasma. Ref.~\cite{Grabowski2013} assumes that the relative velocity between electrons and ions remains approximately constant during a collision, and this is a reasonable approximation for our UNP conditions because the ratio of the electron–ion small impact parameter collision time to RF oscillation period is small ($\omega_0r_L/(2 \pi v_{th}) \ll1$). In this regime, for a VDC the usual $k_{max}$ is multiplied by a factor $1+u^2$~\cite{Grabowski2013}, where $u$ is the ratio of the trajectory particle velocity (the electron in this case) over the electron thermal velocity. In our experiment, this is equivalent to $u=v_{osc}(t)/v_{th}$, where $v_{osc}(t)$ is the instantaneous value of the electron COM velocity at time $t$. For computational convenience, we simplify this by using the average $v_{osc}(t)^2$ weighted by the imparted heating as a function of time. To evaluate the choice of cutoff parameter on predicted RF saturation, we consider two possible cutoff choices within the LRT treatment: one where $k_{max}$ is dependent on the landau length $r_L$, called the velocity-independent cutoff (VIC), and one where $k_{max}$ is dependent on $v_{osc}$ (VDC). 

To test our measurement against these two cutoffs, we write $r_{min}$ as, 
\begin{equation}
 r_{min}=\frac{e^2}{4\pi\epsilon_0m_ev_{eff,i}^2}   
\end{equation}
 where $i$ notes the VIC or VDC treatment. That is,
 \begin{align}
     v_{eff,VIC}&=v_{th} \\
     v_{eff,VDC}^2 &= v_{th}^2+\frac{\int^T_0 v_{osc}^4(t)dt}{\int^T_0 v_{osc}^2(t)dt}
 \end{align}
 for a VIC and VDC respectively. Equation~\ref{eq:bigSVint} is integrated numerically to obtain the heating as a function of RF amplitude.

\subsection{Binary Collision Theory}
While ultimately it does not match our data as well as the LRT with VDC predictions, an approximate BCT provides useful insights into the physics that we observe. In this BCT, the electron--ion collision rate consists of modeling electron--ion collisions as Rutherford scattering~\cite{Rutherford1911} with a maximum impact parameter related to the Debye screening length~\cite{Mulser}. We consider this model for a plasma where the electron velocities $v$ follow a 3D MB distribution. The average heating rate for electrons of the ensemble is then,

\begin{equation}
\begin{aligned}
\frac{dH}{dt}&=
\left({\frac{m_e}{2\pi k_B T_e}} \right)^{3/2}
   \frac{n_e\, v_{osc}(t)\, e^4}{2 \epsilon_0 m_e} \\
&\qquad \times
   \int^{\infty}_0 dv \,
    \exp\left[ -\frac{m_e}{2k_BT_e}\!\left(v^2 + v_{osc}(t)^2\right) \right] \\
&\qquad \times
   \frac{\alpha \cosh(\alpha) -\sinh(\alpha)}{\alpha^2}\,
   \ln\!\left( 1 + \frac{b_{\max}^2}{b_{\perp}^2} \right)
\end{aligned}
\label{eq:BCHeatingRate}
\end{equation}
where $\alpha=m_ev v_{osc}(t)/(k_BT_e)$, $n_e$ is the plasma density, $b_{\perp}$ is a characteristic collision length, and $b_{max}$ is the maximum impact parameter.

While $b_{\perp}$ is often set to be equal to the Landau length $r_L$ in the weakly coupled regime~\cite{Mulser}, its value within this model is $b_{\perp}=b_{\perp,v}=e^2/(4\pi\epsilon_0m_ev^2)$. For weakly coupled plasmas the choice between these two forms of $b_{\perp}$ is not important~\cite{Baalrud2012}. For more strongly coupled plasmas like those described in this work, the two choices produce noticeable differences in predicted heating rates. In BCT, saturation arises because increasing RF intensity increases the electron COM velocity, which increases the relative electron–ion collision speed and reduces the effective collision cross section.

 For our experiments, $v_{th} \sim v_{osc}$ and the plasma frequency $\omega_p$ is similar to the RF frequency $\omega_0$ with $\omega_0=1.8\omega_p$. The UNPs have a spherically symmetric density distribution and so the resonant frequency of the electron COM motion is 0.48$\omega_p$~\cite{Wilson2013}. There is no obvious consensus concerning the form of Coulomb Logarithm (CL) that is best suited to this set of parameters. However, Ref.~\cite{Devriendt2022} performs simulations that align with our highly classical and optics-free conditions. We use that work to determine a reasonable parameterization of the CL for our BCT that can be compared to our experimental measurements. We write the value of $b_{max}$ as
\begin{equation}
    b_{max}^2 = \frac{(a_1v_{th})^2+(a_2v_{osc})^2}{\omega_p^2} 
    \label{eq:CL1} 
\end{equation}
where $a_1$ and $a_2$ are constants to be determined. The constant $a_2$ is associated with a VDC. To set $a_1$ and $a_2$, we compare heating rate predictions as a function of $E_0$ derived from equation (\ref{eq:BCHeatingRate}) against the relative scaling of IBA heating determined from Ref.~\cite{Devriendt2022} for conditions comparable to our experiments while requiring consistency for weakly coupled plasmas in the limit that $E_0 \rightarrow 0$. Doing so produces $a_1=0.7$ and $a_2=0.86$. The fact that $a_2\neq1/6$ is due to the velocity integral in equation (\ref{eq:BCHeatingRate}). For comparison with a VIC, we set $a_2=0$. We compare our experiment to predictions of the heating rate for different assumptions such as the use of $b_{\perp,v}$ vs. $b_{\perp}=r_L$ and VIC vs. VDC. 

For completeness, we also provide a comparison of saturation predictions from BCT consistent with the findings of Ref.~\cite{Turnbull} even though that is not consistent with the parameters derived from Ref.~\cite{Devriendt2022} for our conditions. In this case, the consistent parameters were found to be $a_1=1.34$, $a_2=0$, and an $\omega_0$ dependent $b_{max}$, that is,
\begin{equation}
    b_{max}=b_{max,\omega_0}=\frac{1.34v_{th}}{\omega_0}.
\end{equation}
And the consistent CL is specified in Ref.~\cite{Oster1961} as, 
\begin{equation}
    \ln(b_{max}^2/b_{\perp,v}^2).
\end{equation}

\subsection{Langdon Effect Considerations}

All of the BCT and LRT predictions presented above assume a Maxwell-Boltzmann electron velocity distribution. When $v_{osc}$ is comparable to $v_{th}$, it is predicted that the electron velocity distribution changes away from a Maxwell-Boltzmann distribution~\cite{Langdon1980} in a way that reduces the IBA rate. Although electron--electron collisions tend to restore an MB distribution, their rate is comparable to the electron--ion collision rate and is thus insufficient to fully suppress this effect. For our parameters, a saturation effect of up to 14\% due to the Langdon effect would be expected. This would be in addition to saturation predicted by the theories in the prior two subsections. Such a saturation was not observed, as will be discussed in the results section below.

\subsection{Characteristics of Saturation-Free RF Heating}
In the absence of any saturation effects, the IBA heating will scale with the applied RF intensity and thus linearly with $E_0^2$. To compare the amount of heating from pulses of different lengths and different values of $E_0$, the instantaneous value of the RF electric field $E(t)$ is squared and then integrated for each pulse to obtain a quantity proportional to the amount of heating which can then be used for the basis of comparison.   

\section{Experimental Apparatus and Methods}
\label{sec:ExperimentalApp}

\begin{figure}
    \centering
    \includegraphics[width=\columnwidth]{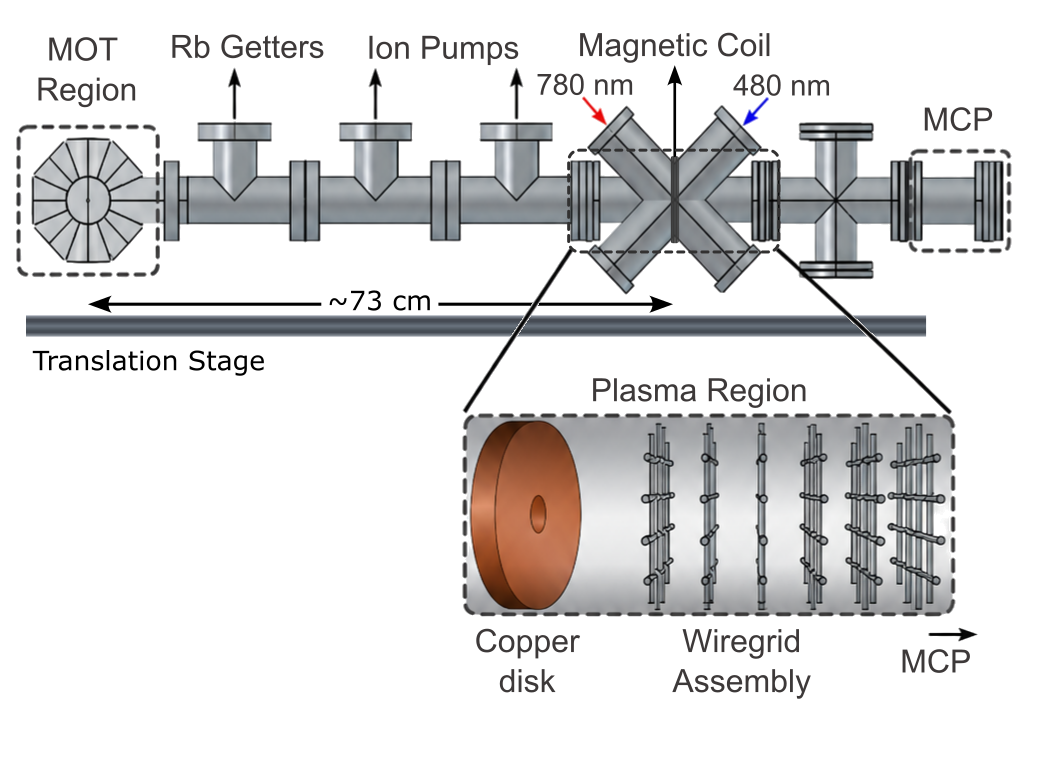}
    \caption{Illustration of our experimental apparatus. $\mathrm{Rb}^{85}$ atoms are laser cooled to $100\,\mu\mathrm{K}$ in a MOT. A motorized magnetic trap transports the atoms to the plasma region where they undergo two-stage photoionization. RF and DC fields are then applied to the electrons through a copper disk and wiregrid. The remaining wiregrids and a magnetic coil then transport the electrons to the MCP detector.}
    \label{fig:ExApp}
\end{figure}

Having described theories whose predictions we compare to our measurements, we next describe our experimental methods. Figure~\ref{fig:ExApp} is given as a reference. 
The first stage of our experimental sequence was trapping and cooling $^{85}\text{Rb}$ atoms in a Magneto-Optic Trap (MOT) to create a gas of atoms with a temperature of about $100\,\mu\text{K}$. The atoms are then transported $\sim73\,\mathrm{cm}$ using a magnetic trap mounted to a motorized translation stage to what we refer to as the ``plasma region". Here, the atoms undergo two-stage photoionization from $780\,\mathrm{nm}$ and $478\,\mathrm{nm}$ lasers~\cite{Killian1999}. The wavelength of the $478\,\text{nm}$ laser was tuned relative to the $^{85}\text{Rb}$ ionization threshold to set the initial electron temperature to $T_e=3.5\,\mathrm{K}$ for the experiments presented. The $780\,\mathrm{nm}$ laser's intensity was adjusted to set the density to $n_e=1.4 \times 10^{13}\,\mathrm{m^{-3}}$. This placed the UNPs in the weakly-to-moderately coupled regime with $\Gamma$ ranging from $0.07-0.19$. DC and RF fields were then applied to the plasma, the details of which will be discussed shortly. A series of wire grids plus a magnetic field were then used to guide electrons that escaped the UNP in response to the applied fields to a microchannel plate (MCP) detector. The plasma had a characteristic cyclotron frequency of $\omega_c=0.89\omega_p$, placing it in the weakly-to-strongly magnetized regime in accordance with the Ref.~\cite{BaalrudDaligault} definition of the weak-to-strong magnetized border as $\omega_c =\omega_p$. 

\begin{figure}[t]
     \centering
     \subfloat[\label{fig:pulsesequences}]{
         \includegraphics[width=\columnwidth]{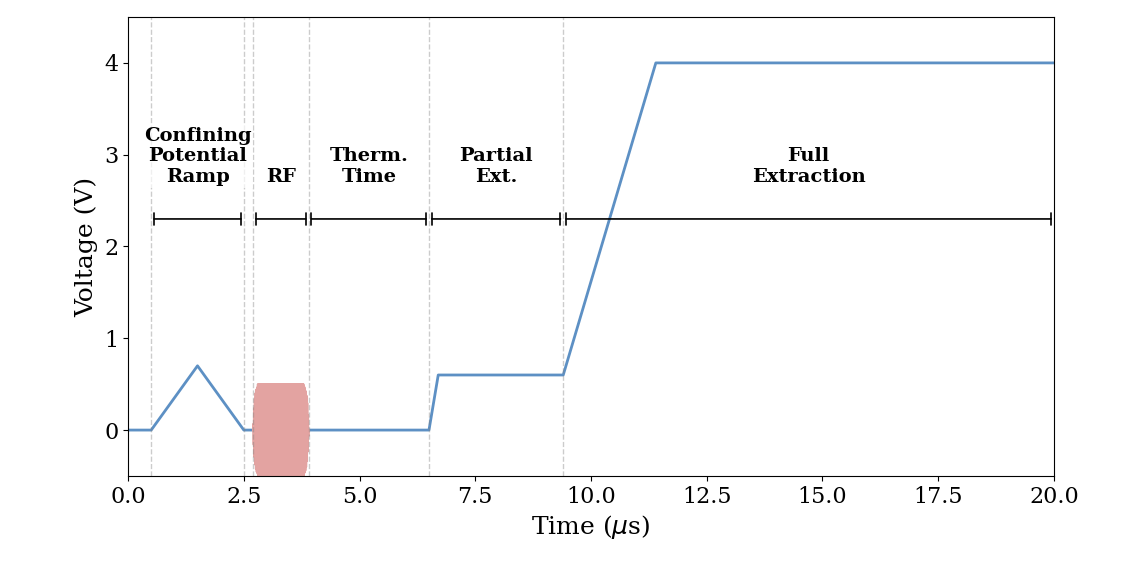}
         \label{fig:appField}
     }
     \\ 
     \subfloat[\label{fig:rawMCP}]{
         \includegraphics[width=\columnwidth]{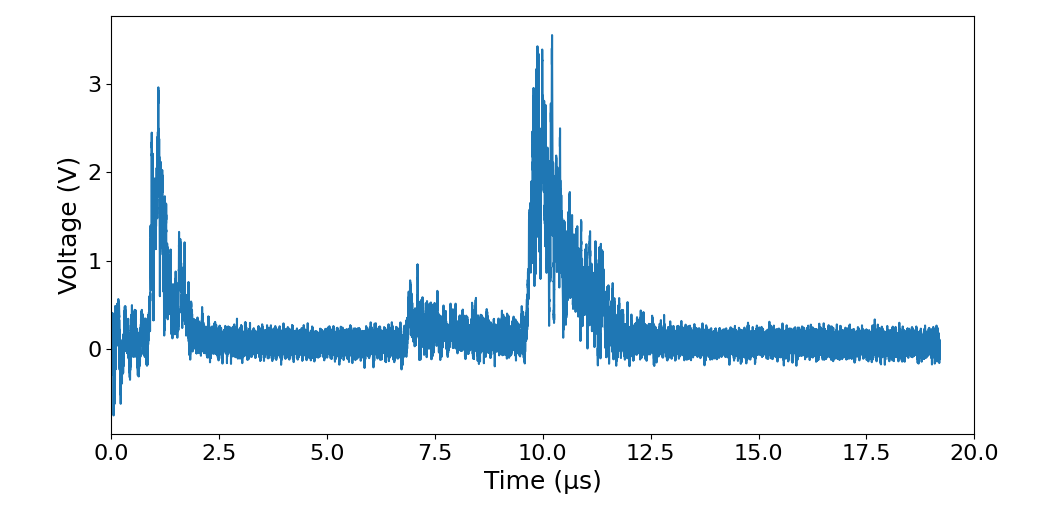}
        \label{fig:MCPtrace}
     }
    
     \caption{(a) DC and RF electric field sequence. (b) Corresponding MCP trace.}
     \label{fig:AppFieldAndMCP}
\end{figure}

After the UNP was formed, we started our experimental sequence as shown in Figure~\ref{fig:appField} with a triangle-shaped ramp of an applied electric field to remove electrons, decreasing the electron to ion number ratio to $70-75\%$ to better confine the remaining electrons. The plasma was then heated via IBA through application of RF fields. The RF electric fields were oriented nearly entirely along the magnetic field direction. Either a short pulse or a long pulse was applied at a frequency of $60\,\text{MHz}$. There were two types of short pulses that were used: a $\tau_p=250\,\text{ns}$ length pulse with a relative voltage amplitude of 1 and a $\tau_p=350\,\text{ns}$ with relative voltage amplitude 0.8. The long pulse was $\tau_p=1200\,\text{ns}$ in length and was applied at relative amplitudes ranging from 0.25 to 0.45. The RF pulses had an adiabatic turn-off and turn-on to isolate IBA heating with a time constant $\tau=33\,\mathrm{ns}$ and amplitude $V_0$. The exact output signal was,
\[
V(t) =
\begin{cases}
(1-e^{-t/\tau})V_0\sin{\omega_0 t} &  t<100\,\text{ns}\\
V_0\sin{\omega_0 t} & 100\,\text{ns}\leq t \le \tau_p-100\,\text{ns}\\
(1-e^{(t-\tau_p)/\tau})V_0\sin{\omega_0 t} &  \tau_p -100\,\text{ns} < t < \tau_p
\end{cases}.
\]
for the various pulses which heated the electrons via IBA. 

While the output of the RF waveform generator are faithful to the functional forms of $V(t)$ shown above, there is a finite response time of the physical elements such as the copper disk and wire grids that are near the UNP. The response of those physical elements is modeled by creating an equivalent RLC circuit and conducting auxiliary experiments to determine the appropriate RLC parameters (see Ref~\cite{GuthrieDens2024} for details). In all theory treatments used, the RF applied to the UNP in the theory calculation includes these finite responses and other effects captured in the RLC description.   

Once the plasma was heated, a delay allowed it to come into thermal equilibrium. An electric field was then applied to extract some of the electrons. The number of electrons extracted was proportional to the plasma heating. We refer to this signal as the partial extraction signal, or PES. Finally, we applied a larger electric field to extract all of the electrons from the plasma for a full extraction that is proportional to the plasma density and compare this to the PES. Figure~\ref{fig:MCPtrace} shows a sample raw MCP trace we would see from 1 experiment sequence.

\begin{figure}[t]
    \centering
    \includegraphics[width=\columnwidth]{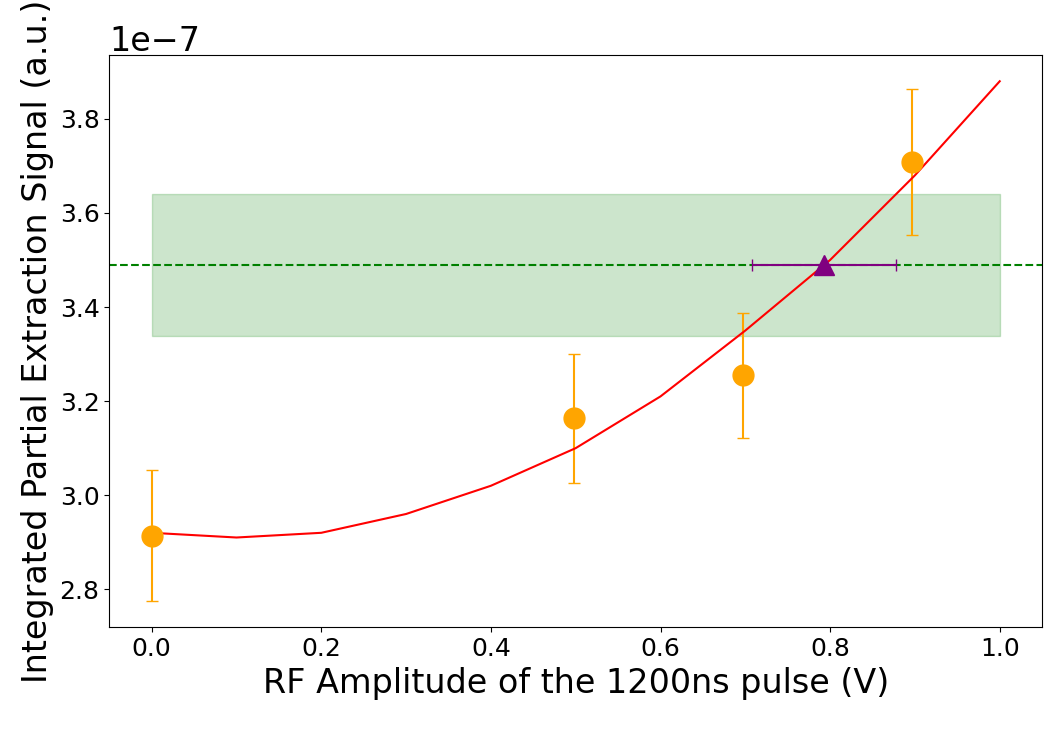}
    \caption{Shown is one of the matched $1200\,\text{ns}$ (purple triangle) pulse amplitudes with error bars to pulse of length $250\,\text{ns}$ (green dashed line). The $250\,\text{ns}$ pulse is fixed at a relative RF amplitude of $1$ and its associated heating is shown by the green dotted line. The green shading around it shows the $250\,\text{ns}$ error range. The orange dots show the $1200\,\text{ns}$ heating for various amplitudes along with the associated error bars. The red line is a $2^{\text{nd}}$ degree polynomial fit to the $1200\,\text{ns}$ data.}
    \label{fig:heatingMatch}
\end{figure}

In the analysis, we found RF amplitudes for different pulses for which the PES is the same, indicating that the same heating had been imparted. A sample matching analysis is given by Figure~\ref{fig:heatingMatch}. In a typical data set, 25-30 individual measurements were made for each RF pulse configuration. One configuration was a short pulse of fixed amplitude. Another three configurations were a long pulse ($1200\,\mathrm{ns}$) of varying amplitudes chosen so that the range of heating from those pulses spanned that from the short pulse. Finally, one other configuration was no RF pulse. By interpolating between the long pulse PES vs. applied RF field, the long-pulse RF amplitude that matched the short-pulse heating was determined. Uncertainties in the PES due to noise (primarily variable background noise from electronic pick-up on the MCP signal from pulsed laser discharge and other sources) translate to uncertainties in the matching amplitude. Multiple data sets were taken to produce precisions in the matching amplitude measurement of better than 3\%.

\begin{figure*}[t]
    \centering

    \subfloat[250 ns pulse\label{fig:250}]{
        \includegraphics[width=0.48\textwidth]{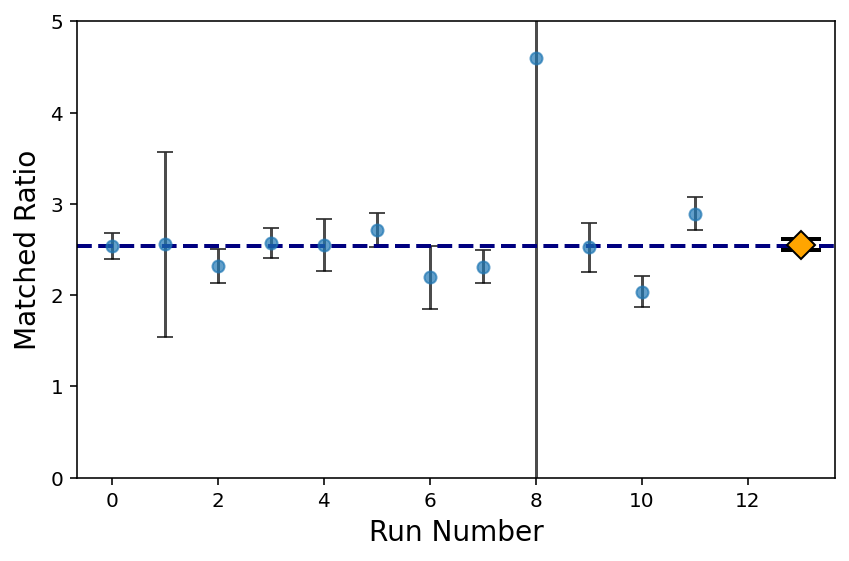}
    }
    \hfill
    \subfloat[350 ns pulse\label{fig:350}]{
        \includegraphics[width=0.48\textwidth]{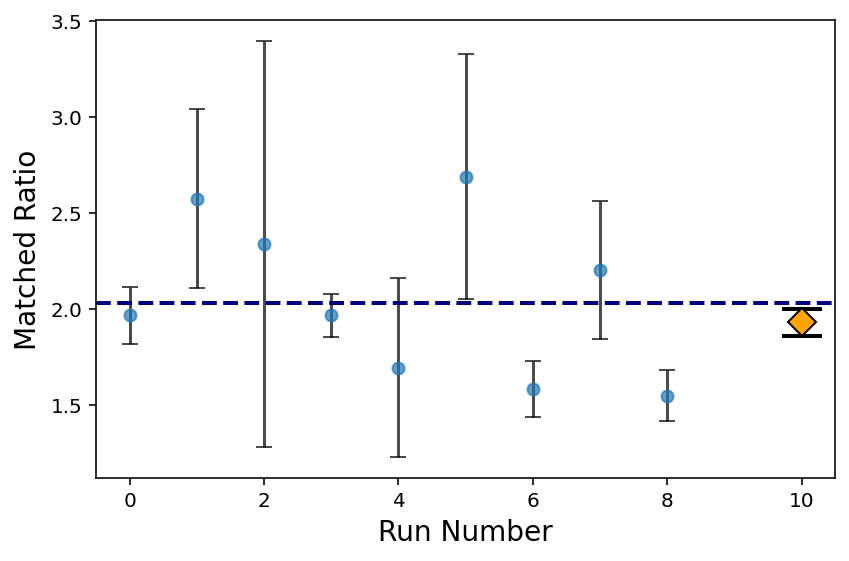}
    }

    \caption{Ratio values for the $250\,\mathrm{ns}$ measurements (a) and the $350\,\mathrm{ns}$ pulse measurements (b) along with their associated errors. The matched ratio is the ratio of short pulse to long pulse RF electric field amplitudes such that the pulses produce the same amount of electron heating. The orange diamonds indicate the average weighted ratios of the $250\,\mathrm{ns}$ (a) and $350\,\mathrm{ns}$ (b) pulses. The dotted lines represent the ratio prediction with a saturation-free model. The y-axis in Figure (a) is truncated so that the averaged error bar can be seen more clearly, although we note that this cuts off one measurement with substantially larger uncertainty.}
    \label{fig:RFsatRatios}
\end{figure*}

There were two calibrations necessary to interpret the data; relating density to laser intensity, and the RF amplitude to a known temperature rise. The density calibration was based on the resonant frequency of the electron COM motion, and details can be found in Ref.~\cite{GuthrieDens2024}. The RF amplitude calibration is discussed here.

To determine the conversion factor for $1200\,\text{ns}$ RF amplitude to electron temperature rise $\Delta T_e$, we measured the PES as a function of applied RF amplitude for $1200\,\text{ns}$ pulses as compared to the increase in PES obtained by increasing the initial UNP electron temperature from $3.5\,\text{K}$ to $6.5\,\text{K}$ via adjusting the photoionization laser wavelength. A matching RF amplitude, which we refer to as $RF_{3K}$, was determined for $3\,\text{K}$ of heating. Changing the photoionization wavelength alters the charge imbalance but by deliberately altering the charge imbalance via slowly applied electric field ramps and interpolation we accounted for that effect~\cite{GuthrieDens2024}. We determined the $1200\,\mathrm{ns}$-pulse RF amplitude that matched $3\,\text{K}$ to a precision of 16\%. 

The main results of the heating comparison measurements are ratios of the amplitudes of the short-duration pulses to matching amplitudes of long-duration pulses that produced the same heating. 

\section{Results and Discussion}
\label{sec:ResultsandDiscussion}

The degree of predicted saturation depends on how much heating is imparted by the RF pulses. To compare the measured ratios with theoretical predictions for a given model (i.e., a specific set of assumptions within LRT or BCT), we introduce a calibration parameter $RF_{3K,\mathrm{model}}$, which is allowed to vary and sets the predicted heating for both short- and long-duration RF pulses. From this, the corresponding predicted electric field amplitude ratios for the $250\,\mathrm{ns}$ and $350\,\mathrm{ns}$ pulses relative to the $1200\,\mathrm{ns}$ pulse are determined. We then perform a best-fit procedure by minimizing $\chi^2$ with respect to the difference between $RF_{3K,\mathrm{model}}$ and $RF_{3K}$ and the difference between predicted and measured pulse amplitude ratios along with the associated uncertainties.

Table~\ref{tab:SummaryTable} summarizes the measured and predicted electric field amplitude ratios for the $250\,\mathrm{ns}$ and $350\,\mathrm{ns}$ short pulses to the $1200\,\mathrm{ns}$ long pulse, along with the $RF_{3K}/RF_{3K,model}$ for each model. Table~\ref{tab:chi_square_stats} shows the corresponding minimum $\chi^2$ values for each model.

\begin{table}[h]
\caption{\raggedright Comparison of measured $250\,\text{ns}$ and $350\,\text{ns}$ to $1200\,\text{ns}$ electric field amplitude ratios with theoretical model predictions, along with the calibrated $RF_{3K}$ value to its theory prediction value $RF_{3K,model}$ ratio for each model. }
\centering
\begin{tabular}{lcccc}
\hline
Model & 250 ns & 350 ns & $RF_{3K}/RF_{3K,model}$ \\
\hline
Measured ratio & 2.55 & 1.93 & -- \\
Measured uncertainty & 0.06 & 0.07 & -- \\
\hline
LRT, VIC & 2.71 & 2.12 & 0.73 \\
LRT, VDC & 2.53 & 1.99 & 1.01 \\
Saturation-free & 2.54 & 2.03 & 1.00 \\
BCT, $b_{\perp,v}$, VDC & 2.60 & 2.05 & 0.88 \\
BCT, $b_{\perp}=r_L$, VDC & 2.63 & 2.07 & 0.82 \\
BCT, $b_{\perp,v}$, VIC & 2.70 & 2.11 & 0.78 \\
BCT, $b_{\perp,v}$, $b_{max,\omega_0}$ (VIC) & 2.70 & 2.06 & 0.78 \\
\hline
\label{tab:SummaryTable}
\end{tabular}

\end{table}

\begin{table}[h]
\caption{\raggedright Error analysis summary, showing $\chi^2$ and $p$-values for each model. }
\centering
\label{tab:chi_square_stats}
\begin{tabular}{lcc}
\hline
Model & $\chi_0^2$ & $P(\chi^2 \ge \chi_0^2)$ \\
\hline
LRT, VIC & 20.00 & $4.54\times10^{-5}$ \\
LRT, VDC & 0.87 & 0.65 \\
Saturation-free & 2.26 & 0.323 \\
BCT, $b_{\perp,v}$, VDC & 4.82 & 0.090 \\
BCT, $b_{\perp}=r_L$, VDC & 8.09 & 0.017 \\
BCT, $b_{\perp,v}$, VIC & 17.02 & $2.01\times10^{-4}$ \\
BCT, $b_{\perp,v}$, $b_{max,\omega_0}$ (VIC) & 16.49 & $2.63\times 10^{-4}$ \\
\hline
\end{tabular}

\end{table}

Interestingly, we observe no measurable saturation of RF heating and no evidence of a Langdon-type distortion of the electron velocity distribution, despite operating in the regime $v_{osc} \sim v_{th}$, where such effects are typically expected. This can be seen in Figure~\ref{fig:RFsatRatios}, which shows the saturation-free model prediction compared to the measured ratio values for both the $250\,\mathrm{ns}$ and $350\,\mathrm{ns}$ pulses, along with the corresponding weighted averaged ratios. This absence of saturation places constraints on models of nonlinear inverse bremsstrahlung heating in moderately coupled classical plasmas.

Both LRT and BCT reproduce the measurements only when VDC is used. In LRT, adopting a VDC for $k_{\max}$, rather than a conventional VIC, yields quantitative agreement with experiment. This supports the general conclusion of Ref.~\cite{Grabowski2013}. Even though that was a prediction for stopping power rather than IBA, for our parameters the predictions of Ref.~\cite{Grabowski2013} are relevant.

\begin{figure}[bt]
    \centering
    \includegraphics[width=\columnwidth]{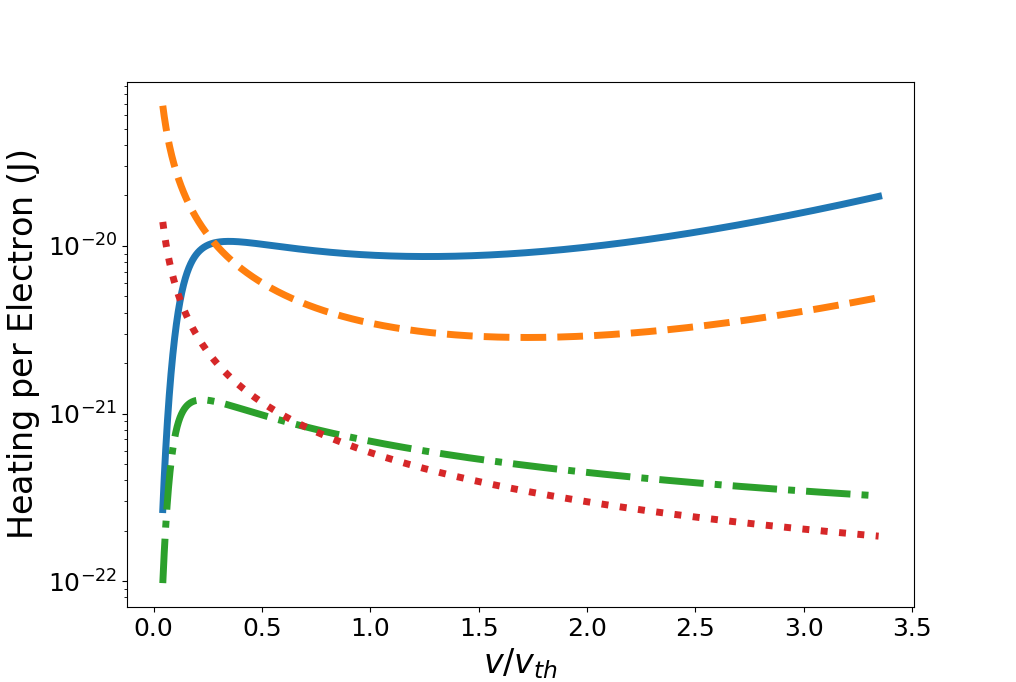}
    \caption{Heating per electron for a MB velocity distribution as a function of $v/v_{th}$. The solid and dashed lines shows the heating for an average Coulomb coupling parameter of $\Gamma=0.185$ ($T_e=3.5\,\text{K}$), while the dotted-dashed and dotted lines show heating for $\Gamma=0.032$ ($T_e=20.0\,\text{K}$). The solid line is for the condition $b_{\perp,v}$ and VDC, while the dashed line shows the heating for $b_{\perp}=r_L$, and VIC. The dotted-dashed and dotted $20.0\,\text{K}$ lines show the heating for the same conditions respectively.}
    \label{fig:HeatingPerElectron}
\end{figure}

The agreement with BCT is more marginal, but improves significantly when both a velocity-dependent $b_{\perp,v}$ and a VDC are included. While this BCT does not provide the same level of quantitative agreement as LRT, it offers useful physical insight. Namely, in the moderately coupled regime, the use of $b_{\perp,v}$ leads to more uniform heating across the electron velocity distribution. This mitigates the preferential heating of low-velocity electrons that underlies the Langdon effect, providing an explanation for its absence under our conditions. To illustrate this further, in Figure~\ref{fig:HeatingPerElectron}, we graph the heating per electron across the MB velocity distribution under different BCT conditions. This graph was generated by taking the heating integrand given by BCT (Equation~\ref{eq:BCHeatingRate}) and dividing it by the MB distribution under different parameters of $\Gamma$, $b_{\perp}$ and $b_{max}$. If the CL is approximated as only dependent on the UNP density and electron temperature (i.e: $b_{\perp}=r_L$), then BCT predicts very different heating rates for low-velocity and high-velocity electrons. However, if the velocity-dependence of the CL is retained ($b_{\perp}=b_{\perp,v}$) then the heating rate for low-velocity electrons is reduced radically as their $b_{\perp}$ value becomes large. When the coupling value is decreased (higher $T_e$), the difference between these two $b_{\perp}$ models becomes small and the heating rate decreases with increased electron velocity $v$. At the higher coupling values we work with (up to $\Gamma=0.19$), the heating is much more uniform across the electron velocity distribution as shown in Figure~\ref{fig:HeatingPerElectron}, reducing the Langdon effect.

Both theoretical approaches do have potentially important limitations in this regime. LRT does not capture large-angle binary collisions, which are expected to become more important at moderate coupling, and does not fully account for the role of collision-induced chaos in the presence of a magnetic field~\cite{Hu2002PhysRevE,Hu2002PoP}. Similarly, while LRT formally includes magnetization, the plasma parameters explored here approach the strongly magnetized regime where existing treatments may be incomplete~\cite{Baalrud2012,BernsteinBaalrud2020}. This version of BCT neglects magnetization effects, treats collisions as instantaneous, and omits recollision dynamics~\cite{Decker1994}. 

More sophisticated BCTs exist and some such as Ref.~\cite{Jose2020} , while associated with stopping power, are consistent with less predicted saturation for our conditions than LRT with VIC. Reproducing predictions from more sophisticated BCTs at the few percent level is beyond the scope of this work. Our measurements provide data that can be used to test such theory predictions.

\section{Conclusions and Future Work}
\label{sec:Conclusions}

In conclusion, we find that our measurements are consistent with no saturation and no observable Langdon effect despite $v_{osc}\sim v_{th}$ in moderately coupled, weakly magnetized, ultracold neutral plasmas. They therefore demonstrate that the onset of nonlinear saturation cannot be inferred solely from the condition that the electron quiver velocity is comparable to the electron thermal velocity. Instead, moderate coupling appears to suppress the Langdon effect, allowing more efficient energy deposition than would be expected based on weakly coupled plasma models. This raises the possibility that similar mechanisms may be relevant in other systems where both strong driving and moderate coupling are present, such as during the picket-pulse time frame in direct-drive ICF experiments~\cite{Craxton2015,Gopalaswamy2019}. Although, there is the possibility that magnetization is playing a more significant role than suggested by linear response theory predictions. Our measurements agree with linear response theory and are most consistent with binary collision theory only when a quiver-velocity-dependent cutoff is used. 

Future work will include measuring saturation at higher electron temperatures where the ultracold neutral plasma is more weakly coupled.
This will enable a bridge between the conditions measured in this work and those more characteristic of IBA in inertial confinement fusion (ICF)-relevant conditions~\cite{Turnbull}.
It is expected that the Langdon effect should become increasingly relevant as the initial electron temperature is increased. We also aim to determine the extent to which the guiding magnetic field can be reduced to test for saturation at lower magnetic fields to decrease the contribution of any affects associated with magnetization. Lastly, we plan to test saturation at even higher effective RF amplitudes by reducing the initial ultracold neutral plasma temperature. We have made some preliminary measurements for the heating as a function of the initial electron temperature. Together, these studies will map out the boundary between linear and nonlinear heating regimes and clarify the conditions under which saturation of inverse bremsstrahlung absorption occurs.

\noindent\textit{Acknowledgments.} This work was supported by the Air Force Office of Scientific Research under Grant No.\ FA9550-21-1-0340.

\noindent\textit{Data availability.} The data that support the findings of this article are available at~\cite{OMara2026}.

\appendix* 
\renewcommand{\thefigure}{A\arabic{figure}}

\clearpage

\bibliographystyle{apsrev4-2}
\bibliography{references}
\end{document}